\documentclass[pre,aps,showpacs,twocolumn]{revtex4}
\usepackage{amsmath}
\usepackage{graphicx}

\begin{document}

\title{Factorization of Dephasing Process in Quantum Open System }
\author{Y. B. Gao}
\email{ybgao@bjut.edu.cn}
\affiliation{College of Applied Science, Beijing University of Technology, Beijing,
100022, China}
\author{C. P. Sun}
\email{suncp@itp.ac.cn}
\homepage{http://www.itp.ac.cn/~suncp}
\affiliation{Institute of Theoretical Physics, Chinese Academy of Sciences, Beijing,
100080, China }

\begin{abstract}
The fluctuation-dissipation relation is well known for the quantum
open system with energy dissipation. In this paper a similar
underlying relation is found between the bath fluctuation and the
dephasing of the quantum open system, of which energy is conserved,
but the information is leaking into the bath. To obtain this
relation we revisit the universal, but simple dephasing model with
quantum non-demolition interaction between the bath and the open
system. Then we show that the decoherence factor describing the
dephasing process is factorized into two parts, to indicate the two
sources of dephasing, the vacuum quantum fluctuation and the thermal
excitations defined in the initial state of finite temperature.
\end{abstract}

\maketitle

\section{Introduction}

A realistic quantum system can rarely be isolated from its surrounding
environment (or called the \textquotedblleft bath") completely \cite%
{Wheeler,open}. When it is coupled to the bath with a large number of
degrees of freedom, decoherence happens. There usually are two distinct
decoherence effects of the bath on the quantum system: when the energy
exchange is allowed by the interaction between the open system and the bath,
the system energy usually dissipates into the environment irreversibly and
we name this effect by quantum dissipation \cite{Leggett, q-disp,Sun95};
another effect is entitled by quantum dephasing, which occurs with no energy
exchange, but an irreversible process of information loss happens in the
considered open system \cite{Zurek91,Privman,q-deco}.

As to quantum dissipation, the well-known fluctuation-dissipation relation
\cite{q-disp, Callen,Fujikawa98,Fujikawa981} reveals to what extent the bath
fluctuation depends on the quantum dissipation process of the open system,
or how the phenomenological damping rate is determined microscopically by
the random couplings to the degrees of freedom of the bath. As to quantum
dephasing, however, there does not exist a similar statement clarified for
the quantum decoherence only with dephasing and without loss of energy.

This paper is devoted to find the intrinsic relation between the pure
dephasing and the some random nature of the bath. We begin with a general
dephasing model for an open system interacting with a bath of many harmonic
oscillators via a coupling of quantum non-demolition \cite{q-m}. This model
is known to be universal in weak coupling limit \cite{Leggett}. Here, we
characterize the dynamic process of dephsing with the so called decoherence
factor, which linearly accompanies the off-diagonal elements of time
evolution of the reduced density matrix for the system obtained by tracing
over the variables of the bath. We find that the decoherence factor is a
product of two factors, one of which is determined by the excitations of the
bath while another does not vanish for the bath initially in vacuum state.
This observation clearly indicates that there exist two sources of quantum
dephasing, which originate from both the vacuum quantum fluctuation and the
thermal excitations of the heat bath respectively.

The paper is organized as follows. In section II, we describe a universal
model for quantum open system interacting with a bath of many bosons through
non-demolition couplings. In section III, we show that a factorization
structure appears in the dynamic dephasing process of the quantum open
system. In section IV, we present the central result of this paper, i.e.,
the relation between dephasing and the fluctuation of the bath. In the
section V, we study the dephasing of the quantum open system while the bath
is in thermal equilibrium state. Finally, in the last section we conclude
this paper with the remarks about the relationship between the quantum
dephasing and the generalized thermalization due to the entanglements of the
system with the bath.

\section{Boson bath for dephasing}

In general, quantum dephasing process is microscopically considered as the
vanishing of off-diagonal elements of the time evolution of reduced density
matrix of an open system interacting with its surrounding environment or
bath of infinite degrees of freedom. The simplest model is the composite
system consisting of a system interacting with the bath of infinite bosons
\cite{Leggett}.

\begin{figure}[th]
\includegraphics[width=5cm]
{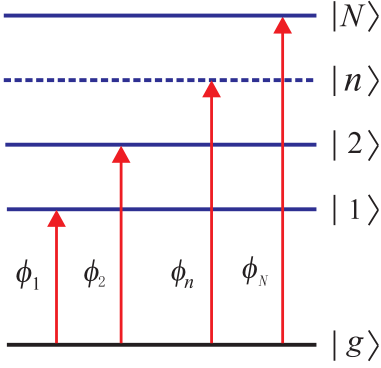}
\caption{The energy level configuration for the quantum open system defined
within the subspace of excited states $|n\rangle $ spanned by the transition
operators $\protect\phi _{n}=|g\rangle \langle n|$ from the ground state $%
\left\vert g\right\rangle $. }
\label{fig-1}
\end{figure}

The model Hamiltonian
\begin{equation*}
H=H_{S}+H_{I}+H_{B}
\end{equation*}%
is decomposed into three parts, the system part $H_{S},$ the bath part $%
H_{B}=\sum_{j}\hbar \omega _{j}a_{j}^{\dag }a_{j}$ with the bosonic creation
(annihilation) operators $a_{j}^{\dag }$ ($a_{j}$) for the mode of
frequencies $\omega _{j}$ ($j=1,2,...$) and the interaction between the
quantum open system and the bath \cite{Sun93}
\begin{equation}
H_{I}=\hbar G\sum\limits_{j}(\xi _{j}a_{j}+\text{H.c.})
\end{equation}%
where we assume that $\hbar \xi _{j}$ has the dimension of energy. The
operator-volumed coupling $G$ depends on the system variables. To conserve
the energy of the open system, if dephasing can happen, the coupling\ is
required to be of quantum non-demolition, i.e., $\left[ H_{S},H_{I}\right]
=0 $ and $\left[ H_{B},H_{I}\right] \neq 0$, or $\left[ H_{S},G\right] =0$.
In general the considered system has $N$ energy levels with eigen-states $%
|n\rangle $ and corresponding eigen-values $\Omega _{n}$ $(n=1,2....N)$ \cite%
{sun-conf}, i.e.,%
\begin{eqnarray}
H_{S} &=&\sum\limits_{n=1}^{N}\hbar \Omega _{n}|n\rangle \langle n|,  \notag
\\
G &=&\sum\limits_{n=1}^{N}g_{n}|n\rangle \langle n|,
\end{eqnarray}%
where $g_{n}$ is dimensionless since $\hbar G\xi _{j}$ has the same
dimension of energy as that of $\hbar \xi _{j}$. The mode number $N$ may be
infinite for a macroscopic heat bath. We assume that the system can be
embedded into a $N+1-$dimensional space with an extended basis vector $%
|g\rangle $ such that $|n\rangle =\phi _{n}^{\dag }|g\rangle $ (see Fig. \ref%
{fig-1}) where the transition operators $\phi _{n}=|g\rangle \langle n|$ has
commutation relations with the projection operators $p_{n}=|n\rangle \langle
n|$
\begin{equation}
\lbrack \phi _{n},p_{m}]=\delta _{mn}\phi _{n}.
\end{equation}

A typical example of this model is that the open system is a single mode
boson with Hamiltonian $H_{S}=\hbar \omega _{0}b^{\dag }b$ defined by the
bosonic creation (annihilation) operator $b^{\dag }$ ($b$) for the mode of
frequency $\omega _{0}$. The interaction $H_{I}=\hbar b^{\dag }b\sum_{j}(\xi
_{j}a_{j}+$H.c.$)$ between the boson and the bath \cite{Zhangp} was
frequently used to give a quantum approach for measurement of boson state.
It originates from the generic oscillator coupling of the system coordinate $%
q$ to the bath variables $x_{j}$, i.e.,

\begin{equation}
\sum_{j}\xi _{j}qx_{j}\sim \sum_{j}[\xi _{j}b^{\dag }a_{j}+\xi _{j}b^{\dag
}a_{j}^{\dagger }+\text{H.c.}]
\end{equation}%
in a large detuning limit $\left\vert \omega _{j}-\omega _{0}\right\vert \gg
\xi _{j}$.

We can exactly solve the Heisenberg equation for the operators $\phi _{n}$
and $a_{j}$%
\begin{align}
\dot{\phi}_{n}& =-i\Omega _{n}\phi _{n}-iXg_{n}\phi _{n},  \notag \\
\dot{a}_{j}& =-i\omega _{j}a_{j}-i\xi _{j}^{\ast }G
\end{align}%
with the quantum noise operator
\begin{equation*}
X=\sum\limits_{j}\left( \xi _{j}a_{j}+\text{H.c.}\right) .
\end{equation*}%
Here, the noise operator satisfies the Brownian conditions in $\left\langle
X(t)\right\rangle =0$ and $\left\langle X(t^{\prime })X(t)\right\rangle \neq
0$ in an equilibrium state $\rho $ and the thermodynamic average $%
\left\langle ...\right\rangle $ is defined as $\left\langle A\right\rangle
=Tr(\rho A).$

Since $G$ is conservable or $G\left( t\right) \mathbf{=}G\left( 0\right)
=G_{0}$, we explicitly obtain
\begin{align}
\phi _{n}(t)& =\phi _{n}(0)e^{-ig_{n}Z(t)+ig_{n}F\left( t\right)
G_{0}}e^{-i\Omega _{n}t}  \notag \\
a_{j}\left( t\right) & =e^{-i\omega _{j}t}a_{j}\left( 0\right) -i\xi
_{j}^{\ast }\eta _{j}\left( t\right) G_{0}  \label{a-operator-time}
\end{align}%
where the phase operator in $b(t)$ reads
\begin{equation*}
Z\left( t\right) =\sum\limits_{j}\left[ \xi _{j}\eta _{j}(t)a_{j}(0)+\text{%
H.c.}\right]
\end{equation*}%
and $\eta _{j}\left( t\right) =i\left( e^{-i\omega _{j}t}-1\right) /\omega
_{j}$. Then the noise operator can be expressed as a linear combination of
initial operators $a_{j}\left( 0\right) $, $a_{j}^{\dag }\left( 0\right) $
and $G_{0}$, i.e.,

\begin{equation}
X(t)=Y(t)-G_{0}\dot{F}\left( t\right)  \label{noise-t}
\end{equation}%
where the \textquotedblleft $q$-number" term
\begin{equation}
Y(t)=\sum\limits_{j}\left[ \xi _{j}e^{-i\omega _{j}t}a_{j}\left( 0\right) +%
\text{H.c.}\right]
\end{equation}%
is due to the free evolution of the reservoir modes, and the $c$-number term
\begin{equation}
F\left( t\right) =2\int \frac{d\omega }{\omega }J\left( \omega \right)
\left( t-\frac{\sin \omega t}{\omega }\right)
\end{equation}%
can be regarded as a time dependent \textquotedblleft force" and defined by
the spectral density function of the bath
\begin{equation}
J\left( \omega \right) =\sum\limits_{j}\left\vert \xi _{j}\right\vert
^{2}\delta \left( \omega -\omega _{j}\right) .  \label{spectral-g}
\end{equation}%
This arises from the back action of the system on the bath.

\section{Factorizing Dephasing Process}

To demonstrate the dynamic process of quantum dephasing of the open system,
we now calculate reduced density matrix for the time evolution of the open
system. A pure dephasing process means that the off-diagonal elements of the
reduced density matrix of the open system vanish, while the diagonal
elements remain unchanged in such an ideal case. Usually the off-diagonal
elements depend on both the initial state and the dynamic variables of the
bath and it can vanish in the thermodynamic limits.

We assume that the initial state of the composite system is of the
factorization form
\begin{equation*}
\left\vert \psi \left( 0\right) \right\rangle =\sum_{n}c_{n}\left\vert
n\right\rangle \otimes \left\vert \left\{ m_{j}\right\} \right\rangle .
\end{equation*}%
It means that the initial state of the open system is a superposition of the
eigen states $\left\vert n\right\rangle =\phi _{n}^{\dag }\left\vert
g\right\rangle $ while the bath is initially in Fock state $\left\vert
\left\{ m_{j}\right\} \right\rangle =\prod_{j}\left\vert m_{j}\right\rangle $%
. Here $m_{j}$ stands for the excitation number of Fock state $\left\vert
m_{j}\right\rangle $. Let $\left\vert 0\right\rangle =\left\vert
g\right\rangle \otimes \left\vert \left\{ 0_{j}\right\} \right\rangle $,
which is invariant under the operation of the evolution operator $U\left(
t\right) =\exp \left( -iHt/\hbar \right) $, i.e., $U\left( t\right)
\left\vert 0\right\rangle =\left\vert 0\right\rangle $.

Then, according to the explicit expressions (\ref{a-operator-time}) for
operators $\phi _{n}$ and $a_{j}$, the time evolution of the composite
system is calculated with a similar method in Ref.\cite{Gao05} as,
\begin{eqnarray*}
\left\vert \psi \left( t\right) \right\rangle &=&U\left( t\right) \left\vert
\psi \left( 0\right) \right\rangle \\
&=&\sum_{n}c_{n}\left[ U\left( t\right) \phi _{n}^{\dag }U^{\dag }\left(
t\right) \right] U\left( t\right) \left\vert g\right\rangle \otimes
\left\vert \left\{ m_{j}\right\} \right\rangle \\
&=&\sum_{n}c_{n}\phi _{n}^{\dag }\left( -t\right) \left\vert g\right\rangle
\otimes \left\vert \left\{ m_{j}\right\} \right\rangle
e^{-i\sum_{j}m_{j}\omega _{j}t} \\
&=&\sum_{n}c_{n}\left( t\right) \left\vert n\right\rangle \otimes
e^{ig_{n}Z(-t)}\left\vert \left\{ m_{j}\right\} \right\rangle
\end{eqnarray*}%
or%
\begin{equation}
\left\vert \psi \left( t\right) \right\rangle =\sum_{n}c_{n}\left( t\right)
\left\vert n\right\rangle \otimes \left\vert \beta _{n}\right\rangle .
\label{state-t-1}
\end{equation}%
Here,%
\begin{equation*}
c_{n}\left( t\right) =c_{n}e^{-i\Omega _{n}t}e^{ig_{n}^{2}F\left( t\right)
}e^{-i\sum_{j}m_{j}\omega _{j}t}
\end{equation*}%
and the coherent state
\begin{equation}
\left\vert \beta _{n}\right\rangle =\prod\limits_{j}D_{j}\left( g_{n}\alpha
_{j}\right) \left\vert m_{j}\right\rangle  \label{coherent-m}
\end{equation}%
is defined by the displacement operators $D_{j}\left( \alpha \right) =\exp
(\alpha a_{j}^{\dag }-\alpha ^{\ast }a_{j})$\ \ with
\begin{equation*}
\alpha _{j}=-i\xi _{j}^{\ast }\eta _{j}\left( t\right) =\frac{\xi _{j}^{\ast
}}{\omega _{j}}\left( e^{-i\omega _{j}t}-1\right) .
\end{equation*}

From the time evolution of density matrix $\rho \left( t\right) =\left\vert
\psi \left( t\right) \right\rangle \left\langle \psi \left( t\right)
\right\vert $ for the composite system. we calculate the reduced density
matrix of the open system
\begin{align}
\rho _{s}\left( t\right) & =\sum_{n}c_{n}c_{n}^{\ast }\left\vert
n\right\rangle \left\langle n\right\vert  \notag \\
& +\sum_{n\neq m}c_{n}c_{m}^{\ast }e^{i\theta _{mn}\left( t\right)
}\left\langle \beta _{m}|\beta _{n}\right\rangle \left\vert n\right\rangle
\left\langle m\right\vert
\end{align}%
by tracing over the variables of the bath. Here%
\begin{equation*}
\theta _{mn}\left( t\right) =\left( \Omega _{m}-\Omega _{n}\right) t+\left(
g_{n}^{2}-g_{m}^{2}\right) F\left( t\right)
\end{equation*}%
is the time dependent real number.

To characterize the coherence of the quantum open system, we use the
decoherence factor \cite{Sun93}%
\begin{equation}
D_{n,m}\left( t\right) =\left\langle \beta _{m}|\beta _{n}\right\rangle
\equiv \prod\limits_{j}D_{mn}[m_{j}].  \label{dfactor-1}
\end{equation}%
Initially, the decoherence factor $D_{n,m}\left( 0\right) =1$, i.e., the
system owns a completely ideal quantum coherence. Here, each of two factors
in the decoherence factor
\begin{equation*}
D_{mn}[m_{j}]=\left\langle m_{j}\right\vert D_{j}\left( \left(
g_{n}-g_{m}\right) \alpha _{j}\right) \left\vert m_{j}\right\rangle ,
\end{equation*}%
is calculated as
\begin{equation}
D_{mn}\left[ m_{j}\right] =e^{-\frac{1}{2}z_{nm;j}(t)}L_{m_{j}}\left(
z_{nm;j}(t)\right)
\end{equation}%
in terms of the Laguerre polynomial $L_{m_{j}}\left( z_{nm;j}(t)\right) $ of
variable
\begin{equation*}
z_{nm;j}(t)=\left( g_{n}-g_{m}\right) ^{2}\left\vert \xi _{j}\eta _{j}\left(
t\right) \right\vert ^{2}
\end{equation*}

With the above results, we can observe that the decoherence factor can be
decomposed into two parts, i.e.,
\begin{equation}
D_{n,m}\left( t\right) =D_{n,m}^{\left( 0\right) }\left( t\right)
D_{n,m}^{\left( \left\{ m_{j}\right\} \right) }\left( t\right)
\label{dfactor}
\end{equation}%
where
\begin{equation*}
D_{n,m}^{\left( 0\right) }\left( t\right) =\prod\limits_{j}e^{-\frac{1}{2}%
z_{nm;j}(t)}
\end{equation*}%
originates from the vacuum quantum fluctuation of \ the bath and
\begin{equation}
D_{n,m}^{\left( \left\{ m_{j}\right\} \right) }\left( t\right)
=\prod\limits_{j}L_{m_{j}}\left( z_{nm;j}(t)\right)  \label{d-l}
\end{equation}%
means the bath excitation. Actually, when the bath is initially in the
vacuum state $\left\vert \left\{ m_{j}=0\right\} \right\rangle $, the second
factor $D_{n,m}^{\left( \left\{ m_{j}\right\} \right) }\left( t\right) $
becomes a unity operator since $L_{m_{j}=0}\left( x\right) =1.$ The
decoherence factor $D_{n,m}\left( t\right) =D_{n,m}^{\left( 0\right) }\left(
t\right) $ means that the dephasing only results from the vacuum fluctuation
of the bath without thermal excitation. When the initial state of the bath $%
\left( \left\vert \left\{ m_{j}\right\} \right\rangle \right) $ is occupied
by large amount of excitation, the macroscopic feature of the bath (high
excitations) can induce the dephasing of the open system. This is because $%
L_{m_{j}}\left( x\right) $ approaches the zero-order Bessel function $%
J_{0}\left( x\right) $ when $m_{j}\rightarrow \infty $ \cite{Sun97}. Then%
\begin{equation}
D_{mn}\left[ m_{j}\right] =e^{-\frac{1}{2}z_{nm;j}(t)}J_{0}\left(
z_{nm;j}(t)\right) .
\end{equation}%
Since the Bessel function with real variables is a decaying oscillating
function, the decoherence factor $D_{n,m}\left( t\right) $ (\ref{dfactor})
approaches zero when $t$ tends to infinity. We will discuss how a thermal
equilibrium state can induce the fast dephasing. The above arguments show
that, even though the system energy is conserved, there still exist the two
sources of the pure quantum dephasing induced by the bath, i.e., the vacuum
quantum fluctuation and the thermal excitations, which leak the information
of the system into the bath.

\begin{figure}[th]
\includegraphics[width=6cm]{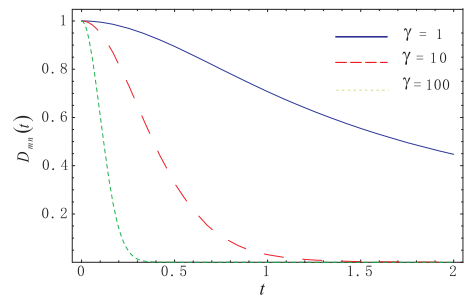}
\caption{Plot of $D_{n,m}\left( t\right) $ (decoherence factor) (\protect\ref%
{dd}) vs $t$ (time) for various value of the coupling parameter $\protect%
\gamma $. $\left\vert g_{m}-g_{n}\right\vert =1$ and $\Gamma =1$.}
\label{d}
\end{figure}

We can explicitly demonstrate these results in the continuous limit. As a
illustration we take Ohmic spectral density of the bath \cite{Leggett} \cite%
{Allahverdyan04}
\begin{equation}
J\left( \omega \right) =\gamma \omega e^{-\omega /\Gamma }
\end{equation}%
to make the sum in
\begin{equation}
D_{n,m}^{\left( 0\right) }\left( t\right) =e^{-\left( g_{n}-g_{m}\right)
^{2}\sum_{j}\left\vert \xi _{j}\eta _{j}\left( t\right) \right\vert ^{2}/2}
\end{equation}%
as a \ integral
\begin{equation}
\sum\limits_{j}\left\vert \xi _{j}\eta _{j}(t)\right\vert ^{2}=\int d\omega
J\left( \omega \right) \frac{4}{\omega ^{2}}\sin ^{2}\frac{\omega t}{2}.
\end{equation}%
Here, $\gamma $ is a dimensionless coupling constant, and $\Gamma $ the
bath's response frequency. When the initial state of the bath is in vacuum
state, the decoherence factor in Eq.(\ref{dfactor}) becomes%
\begin{equation}
D_{n,m}\left( t\right) =\left( 1+\Gamma ^{2}t^{2}\right) ^{-\frac{1}{2}%
\left( g_{n}-g_{m}\right) ^{2}\gamma }.  \label{dd}
\end{equation}%
Figure \ref{d} shows that the decoherence factor (\ref{dd}), the decay of
which is similar to an exponential decay. Actually, at short time limit, $%
\Gamma ^{2}t^{2}\ll 1$, the decoherence factor (\ref{dd}) becomes Gaussian
decaying faction as
\begin{equation}
D_{n,m}\left( t\right) =e^{-\left( t/\tau \right) ^{2}}.
\end{equation}%
Here the characteristic time $\tau $ of the dephasing of the open system is
defined by
\begin{equation*}
\tau ^{-1}=\left( g_{n}-g_{m}\right) \Gamma \sqrt{\frac{\gamma }{2}}.
\end{equation*}

\section{Dephasing-Fluctuation Relation}

Now we take another approach to calculate the decoherence factor, which will
offer us a new angle to understand the source of dephasing. Here, we adopt
the notation of the standard deviation $\Delta A=\sqrt{\left\langle
A^{2}\right\rangle -\left\langle A\right\rangle ^{2}}$ for a given operator $%
A$. We write down the decoherence factor
\begin{equation}
D_{n,m}\left( t\right) =\left\langle \left\{ I_{j}\right\} \right\vert
e^{i\phi \left( t\right) }\left\vert \left\{ I_{j}\right\} \right\rangle
\end{equation}%
in terms of the phase difference operator%
\begin{equation*}
\phi \left( t\right) =\left( g_{n}-g_{m}\right) Z(-t)
\end{equation*}%
for the bath initially with a factorized state $\left\vert \left\{
I_{j}\right\} \right\rangle =\prod_{j}\left\vert I_{j}\right\rangle $. We
denote the average $\langle \left\{ I_{j}\right\} \left\vert A|\left\{
I_{j}\right\} \right\rangle $ by$\ \langle A\rangle .$

For the small variation $|\phi \left( t\right) -\left\langle \phi
(t)\right\rangle |$ of the phase $\phi \left( t\right) $ around it average $%
\left\langle \phi (t)\right\rangle $ over any state we have approximately
\cite{Aharonov}%
\begin{equation}
D_{n,m}\left( t\right) =\langle e^{i\phi \left( t\right) }\rangle \simeq
e^{i\left\langle \phi (t)\right\rangle }e^{-\frac{1}{2}\left( \Delta \phi
(t)\right) ^{2}}.
\end{equation}%
It shows that the phase fluctuation $\Delta \phi (t)$ results in the loss of
quantum coherence or quantum dephasing characterized by the decaying of the
decoherence factor. For the initial Fock state with $\left\vert
I_{j}\right\rangle =\left\vert m_{j}\right\rangle $ we discussed in the last
section, $\left\langle \phi (t)\right\rangle =0$ and the phase fluctuation
due to the bath fluctuation can be separated explicitly into two parts,
i.e.,
\begin{equation}
\left( \Delta \phi (t)\right) ^{2}=\left( \Delta \phi (t)\right)
_{0}^{2}+\left( \Delta \phi (t)\right) _{f}^{2}.
\end{equation}%
where the vacuum fluctuation part is
\begin{equation}
\left( \Delta \phi (t)\right) _{0}^{2}=\sum_{j}z_{nm;j}(t)
\end{equation}%
and the bath excitation part
\begin{equation}
\left( \Delta \phi (t)\right) _{f}^{2}=\sum_{j}2m_{j}z_{nm;j}(t).
\end{equation}

Actually the result is the same as what we have obtained above in some
cases. This is because the couplings are linear with respect to the
operators $a_{j}$ and $a_{j}^{\dag }$. Through some simple calculations, we
obtain the decoherence factor as
\begin{eqnarray}
D_{n,m}\left( t\right) &=&e^{-\frac{1}{2}\left( \Delta \phi (t)\right)
_{0}^{2}}e^{-\frac{1}{2}\left( \Delta \phi (t)\right) _{f}^{2}}  \notag \\
&\equiv &D_{n,m}^{\left( 0\right) }\left( t\right) D_{n,m}^{\left( f\right)
}\left( t\right)  \label{dfactor-js}
\end{eqnarray}%
where the thermal excitation part can be rewritten as
\begin{equation}
D_{n,m}^{\left( f\right) }\left( t\right) =\prod\limits_{j}e^{-f_{j}\left(
m_{j}\right) }=\prod\limits_{j}e^{-m_{j}z_{nm;j}(t)}.
\end{equation}%
When the bath is initially in vacuum state $\left\vert \left\{ m_{j}\right\}
\right\rangle =\left\vert \left\{ m_{j}=0\right\} \right\rangle $, $%
f_{j}\left( m_{j}\right) =0$. Then the decoherence factor becomes $%
D_{n,m}\left( t\right) =D_{n,m}^{\left( 0\right) }\left( t\right) $.
Otherwise we have $f_{j}\left( m_{j}\right) >0$, it means that the
decoherence factor in Eq.(\ref{dfactor-js}) will decay with time .

Next we consider the difference of the decoherence factors obtained through
different approaches in Eq.(\ref{dfactor}) and Eq.(\ref{dfactor-js})
respectively. When $m_{j}$ and $\left\vert \left( g_{n}-g_{m}\right) \xi
_{j}\right\vert /\omega _{j}$ are all small, i.e., the bath is setup in a
low-excitation state and coupling is weak, the Laguerre polynomial in Eq.(%
\ref{d-l}) can be approximately as%
\begin{equation}
L_{m_{j}}\left( x\right) \simeq e^{-m_{j}x}.
\end{equation}%
Then the two decoherence factor in Eq.(\ref{dfactor}) and Eq.(\ref%
{dfactor-js}) has the same form. We consider the time evolution of the
average excitation number of the bath, i.e.,%
\begin{equation}
N_{B}\left( t\right) =\sum_{j}\left\langle a_{j}^{\dag }\left( t\right)
a_{j}\left( t\right) \right\rangle =N_{B}\left( 0\right) +\delta N_{B}\left(
t\right)  \label{mean number}
\end{equation}%
where the initial average excitation number of the bath is $N_{B}\left(
0\right) =\sum_{j}m_{j}$, and
\begin{align}
\delta N_{B}\left( t\right) & =\left\langle G_{0}^{2}\right\rangle
\sum_{j}\left\vert \xi _{j}\eta _{j}\left( t\right) \right\vert ^{2}  \notag
\\
& =\left\langle G_{0}^{2}\right\rangle \int d\omega J\left( \omega \right)
\frac{4}{\omega ^{2}}\sin ^{2}\frac{\omega t}{2}  \label{fluctuation}
\end{align}%
is the quantum fluctuation of the bath excitation numbers, which is
independent of the initial state of the bath. In the above equation, the
term $\left\langle G_{0}^{2}\right\rangle $ indicates the average of the
square of the population number of the open system. This means
\begin{equation}
\sum\limits_{j}\left\vert \xi _{j}\eta _{j}\left( t\right) \right\vert ^{2}=%
\frac{\delta N_{B}\left( t\right) }{\left\langle G_{0}^{2}\right\rangle }.
\end{equation}%
Then the decoherence factor (\ref{dfactor}) can be rewritten in terms of the
fluctuation of excitation number,
\begin{equation}
D_{n,m}\left( t\right) \propto \exp \left[ -\frac{1}{2}\left(
g_{n}-g_{m}\right) ^{2}\frac{\delta N_{B}\left( t\right) }{\left\langle
G_{0}^{2}\right\rangle }\right]
\end{equation}

This equation shows that the decoherence factor is determined by the quantum
fluctuation of bath $\delta N_{B}\left( t\right) $ and $\left\langle
G_{0}^{2}\right\rangle $. When the initial state of the open system is
given, the value of $\left\langle G_{0}^{2}\right\rangle $ is fixed in the
mean while. That is to say, $\left\langle G_{0}^{2}\right\rangle $ is not
relevant to the source of dephasing of the open system, but the energy of
the open system will contribute to the dephasing. Thus we conclude that the
quantum fluctuation of the bath excitation characterized by $\delta
N_{B}\left( t\right) $ induces the dephasing of the open system.

\section{Dephasing in Thermal Equilibrium State}

When the bath is prepared in a pure state we have found that the
dephasing-fluctuation relation for quantum dephasing is similar to the
well-known fluctuation-dissipation relation for quantum dissipation.
Correspondingly the dephasing process can be understood separately according
to the quantum fluctuation and the pure state excitation of the bath. In
this section we will show that these observations also hold exactly for the
case that the bath is initially in a thermal equilibrium state.

For the dephasing problem of the open system at finite temperature, the
thermal equilibrium bath is described by the density matrix%
\begin{equation}
\rho _{B}=\prod\limits_{j}\frac{e^{-\beta \hbar \omega _{j}a_{j}^{\dag
}a_{j}}}{Tr\left( e^{-\beta \hbar \omega _{j}a_{j}^{\dag }a_{j}}\right) }.
\end{equation}%
Initially, the density operator of the composite system is a direct product
\begin{equation*}
\rho \left( 0\right) =\left\vert \psi \left( 0\right) \right\rangle
\left\langle \psi \left( 0\right) \right\vert \otimes \rho _{B}
\end{equation*}
with the initial state $\left\vert \psi \left( 0\right) \right\rangle
=\sum\nolimits_{n}c_{n}\left\vert n\right\rangle $ of the open system. Using
the coherent-state representation, the density operator can be rewritten as
\begin{equation}
\rho _{B}=\prod\limits_{j}\int d^{2}\lambda _{j}\rho \left( \lambda
_{j}\right) \left\vert \lambda _{j}\right\rangle \left\langle \lambda
_{j}\right\vert
\end{equation}%
with the $P-$representation for diagonal elements%
\begin{equation}
\rho \left( \lambda _{j}\right) =\frac{1}{\pi \left\langle
m_{j}\right\rangle }e^{-\left\vert \lambda _{j}\right\vert ^{2}/\left\langle
m_{j}\right\rangle }
\end{equation}%
where $\left\langle m_{j}\right\rangle =\left( e^{\beta \hbar m_{j}\omega
_{j}}-1\right) ^{-1}$ is the average excitation number in the mode of the
frequency $\omega _{j}$.

By applying the results obtained in the previous sections, we can calculate
the reduced density matrix
\begin{equation}
\rho _{s}\left( t\right) =\sum_{n}c_{n}c_{n}^{\ast }\left\vert
n\right\rangle \left\langle n\right\vert +\sum_{n\neq m}c_{n}\left( t\right)
c_{m}^{\ast }\left( t\right) \left\vert n\right\rangle \left\langle
m\right\vert D_{n,m}^{\left[ T\right] }\left( t\right) .
\end{equation}%
to characterize the quantum coherence of the open system. Then the
decoherence factor in the above equation is factorized as \cite{Sun97}
\begin{align}
D_{n,m}^{\left[ T\right] }\left( t\right) & =\prod\limits_{j}\int
d^{2}\lambda _{j}\frac{e^{-\left\vert \lambda _{j}\right\vert
^{2}/\left\langle m_{j}\right\rangle }}{\pi \left\langle m_{j}\right\rangle }%
\left\langle \lambda _{j}\right\vert D_{j}\left( \alpha _{jmn}\right)
\left\vert \lambda _{j}\right\rangle  \notag \\
& =D_{n,m}^{\left( 0\right) }\left( t\right) \prod\limits_{j}e^{-g_{j}\left(
T\right) }  \label{d-tem}
\end{align}%
where $\alpha _{jmn}=\left( g_{n}-g_{m}\right) \alpha _{j}$ and
\begin{equation}
g_{j}\left( T\right) =\frac{z_{nm;j}}{e^{\beta \hbar \omega _{j}}-1}.
\end{equation}

\begin{figure}[th]
\includegraphics[width=6cm]{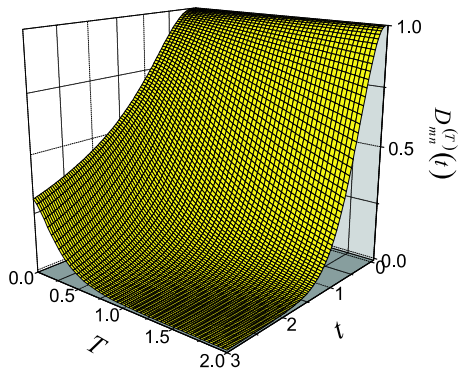}
\caption{Plot of $D_{n,m}^{\left[ T\right] }\left( t\right) $ (decoherence
factor) (\protect\ref{d-tem}) vs $t$ (time) and $T$ (temperature). $%
\left\vert g_{m}-g_{n}\right\vert =1$, $\protect\gamma =1$ and $\Gamma =1$.}
\label{fig-tem}
\end{figure}
Figure \ref{fig-tem} shows that the decoherence factor $D_{n,m}^{\left[ T%
\right] }\left( t\right) $ depends on the time $t$ and the temperature $T$
of the bath. When the temperature $T$ increases, the decaying decoherence
factor will be enhanced, i.e., the open system loses its quantum coherence.

When the temperature approaches the absolute zero degree ($\beta \rightarrow
\infty $), we approximately have $\left( e^{\beta \hbar \omega
_{j}}-1\right) ^{-1}\simeq e^{-\beta \hbar \omega _{j}}$. Then the
decoherence factor (\ref{d-tem}) becomes%
\begin{equation}
D_{n,m}^{\left[ T\right] }\left( t\right) =D_{n,m}^{\left( 0\right) }\left(
t\right) \exp \left( -\frac{\left( g_{n}-g_{m}\right) ^{2}\gamma t^{2}}{%
\hbar ^{2}}k_{B}^{2}T^{2}\right) .
\end{equation}%
It shows that, in the low-temperature limit, the above decoherence factor
exponentially decays as the square of temperature $T^{2}$ increases, which
is called Gaussian decay. When the bath is prepared in the high temperature (%
$\beta \hbar \omega _{j}\rightarrow 0$), we have $e^{\beta \hbar \omega
_{j}}-1\simeq \beta \hbar \omega _{j}$, and then%
\begin{equation}
g_{j}\left( T\right) =\left( g_{n}-g_{m}\right) ^{2}\frac{\left\vert \xi
_{j}\eta _{j}\left( t\right) \right\vert ^{2}}{\beta \hbar \omega _{j}}%
\propto T.
\end{equation}%
It means that, when the bath approaches the high-temperature limit, the
decoherence factor (\ref{d-tem}) exponentially decays as the temperature $T$
increases.

\begin{figure}[th]
\includegraphics[width=6cm]{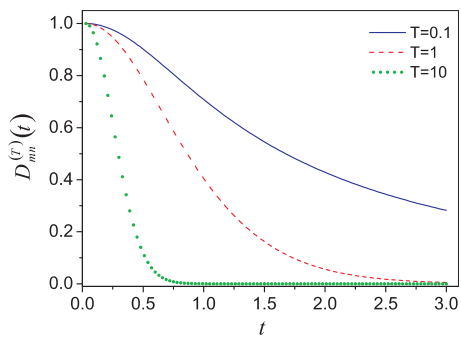}
\caption{Plot of $D_{n,m}^{\left[ T\right] }\left( t\right) $ (decoherence
factor) (\protect\ref{d-tem}) vs $t$ (time) for various value of temperature
$T$. $\left\vert g_{m}-g_{n}\right\vert =1$, $\protect\gamma =1$ and $\Gamma
=1$.}
\label{dfg}
\end{figure}
In addition, Figure \ref{dfg} shows that the rise of temperature will
accelerate dephasing described by Eq.(\ref{d-tem}).

\section{Conclusion with remarks}

In summary, through a universal model for a quantum open system with
non-demolition coupling to the bath, we find the intrinsic relation between
the dephasing of the open system and the quantum and thermal fluctuation of
its bath. Usually, the couplings of a quantum open system to its environment
is very complicated, and then intuitively the dephasing process should
depend on the details of system-bath couplings. So generally the present
model used in this paper seems to be too oversimplified. However, we would
like to mention the Caldeira and Leggett's studies about quantum dissipation
\cite{Leggett}, which proves that any bath with weak couplings to the open
system can be modeled as a collection of non-interacting bosons. As for the
quantum dephasing, we even found a similar conclusion to universally
consider the quantum dephasing problem in quantum computing \cite%
{Sun-Zhan-Liu}. Thus the present investigation can also be considered to be
universal, and dephasing process is generally clarified as the two kinds of
origins, the quantum fluctuation in the vacuum and the thermal excitation in
finite temperature.

Before concluding this paper, it is necessary to say some words about the
repletion and difference between the dynamic dephasing process studied in
this paper and the thermalization to the thermal equilibrium induced by the
bath \cite{gt-nature, gt-typic, gt-tasaki, gz-m, gt-loyd}. First, we have to
note that the dephasing process in this paper can describe the thermal
equilibrium with canonical state in a straightforward way. Most recently
people revisit the investigation to\emph{\ }explore the possibility
replacing the equal a prior probability postulate in statistical mechanics
by a general canonical principle \cite{gt-nature}: an arbitrary entangling
pure state of the total system consisting of the \textquotedblleft small"
system plus \textquotedblleft large bath" can be traced over the variables
of the bath to give a generalized canonical state due to the
\textquotedblleft Large Number Law" or overwhelming majority of wave
functions in the subspace by obeying some global constraint for the
\textquotedblleft universe". Actually, if this global constraint is
particularly determined by the energy interval encompassed by the
microcanonical ensemble, the above mentioned \textquotedblleft tracing"
operation just results in the thermal equilibrium distribution. Such
thermalization process can also be described as a typical quantum
dissipation phenomenon, a dynamically quantum process with energy exchange
between the system and the bath \cite{ozg,open}. However, there is not
energy exchange in our present dephasing model and thus it give rise to a
mixture rather than the thermal equilibrium distribution. Maybe there exist
another type global constraint other than the energy interval, we believe
that it is an open question, which could be solved in a general framework.
We will continue the exploration along this line in future studies.

\begin{acknowledgments}
We acknowledge the support of the NSFC (grant No. 90203018, 10474104,
10447133), the Knowledge Innovation Program (KIP) of Chinese Academy of
Sciences, the National Fundamental Research Program of China (No.
2001CB309310). Y. B. Gao is also supported by the NSFC (grant No. 10547101)
and the Science Funds for PhD. in Beijing University of Technology. And we
would like to thank J. Lu for his help in drawing the pictures in this paper.
\end{acknowledgments}

\end{document}